\documentclass[twocolumn,twocolappendix]{aastex631}

\usepackage{comment}
\accepted{August 12, 2025}

\begin{document}

\title{Halo Occupation Distribution of Quasars: Dependence on Luminosity, Redshift, Black Hole Mass and Feedback Modes}

\author[0009-0004-1967-0474]{Anirban Chowdhary}
\affiliation{School of Astrophysics, Presidency University\\
86/1, College Street, \\Kolkata, India}
\email{anirban.rs@presiuniv.ac.in}

\author[0000-0002-3236-2853]{Suchetana Chatterjee}
\affiliation{School of Astrophysics, Presidency University\\
86/1, College Street, \\Kolkata, India}
\email{suchetana.astro@presiuniv.ac.in}

\begin{abstract}

 We use cosmological hydrodynamic simulations (IllustrisTNG and SIMBA) to explore the redshift, luminosity, and black hole mass dependence of the quasar halo occupation distribution (HOD). In both simulations, we find that the quasar activity is quenched at a characteristic halo mass ($\sim 10^{13} M_{\odot}$) scale affecting the nature of its occupation distribution function. We note that the quenching is more pronounced at low redshifts for quasars selected through a luminosity threshold. We show that a very significant bias (a factor of $\sim 10-50$ in the central occupation and $\sim 10-70\%$ in the satellite occupation fraction) is introduced in the reconstruction of quasar host halo mass distributions from the observed two-point-correlation function, if the HOD modeling does not account for the quenching effect in the central occupation function. While there is strong suppression of the occupation fraction of central quasars, the satellite occupation still follows a power-law like behavior. Our results show that the global satellite fraction of quasars increases monotonically from high to low redshifts, with $20-40 \%$ of the quasars being satellite at intermediate redshifts, consistent with previous clustering based estimates. In addition, our study reveals that while the occupation function of quasars depends on redshift, luminosity, and feedback modes, there is hardly any evolution in the supermassive black hole (SMBH; mass-selected sample) occupation. The SMBH HOD over the entire parameter space is well-modeled by a power-law and a step function similar to what has been found for galaxies and low-luminosity active galactic nuclei.

 
 \end{abstract}


\section{Introduction} \label{sec:Intro}
In the $\Lambda$CDM paradigm, baryonic matter falls into the potential well of the dark matter (DM) halos forming galaxies \citep[e.g.,][]{w&r78, w&f91, kauffmannetal93, nfw95, m&w96, kauffmannetal99, hopkinsetal10, c&w13, conselice14, shankaretal15, Yang&Mo&Bosch03,Yang&Mo&Bosch08,Yang12,Behroozietal13}. The galaxies themselves harbor supermassive black holes (SMBH) at their centers. Numerous observational evidences suggest that SMBH greatly influence the evolution of their host galaxies (see \citealt{Kormendy&Ho13} for a review). Hence, it is evident that there exists an intricate connection linking SMBH with their host galaxies and subsequently the halo potential that they are part of. A complete knowledge of galaxy evolution thus requires an understanding of the relation between the growth of SMBH and the DM halos that they inhabit, in addition to the host galaxy-SMBH connection. 

\begin{figure*}[t] 
            \includegraphics[width=\textwidth]{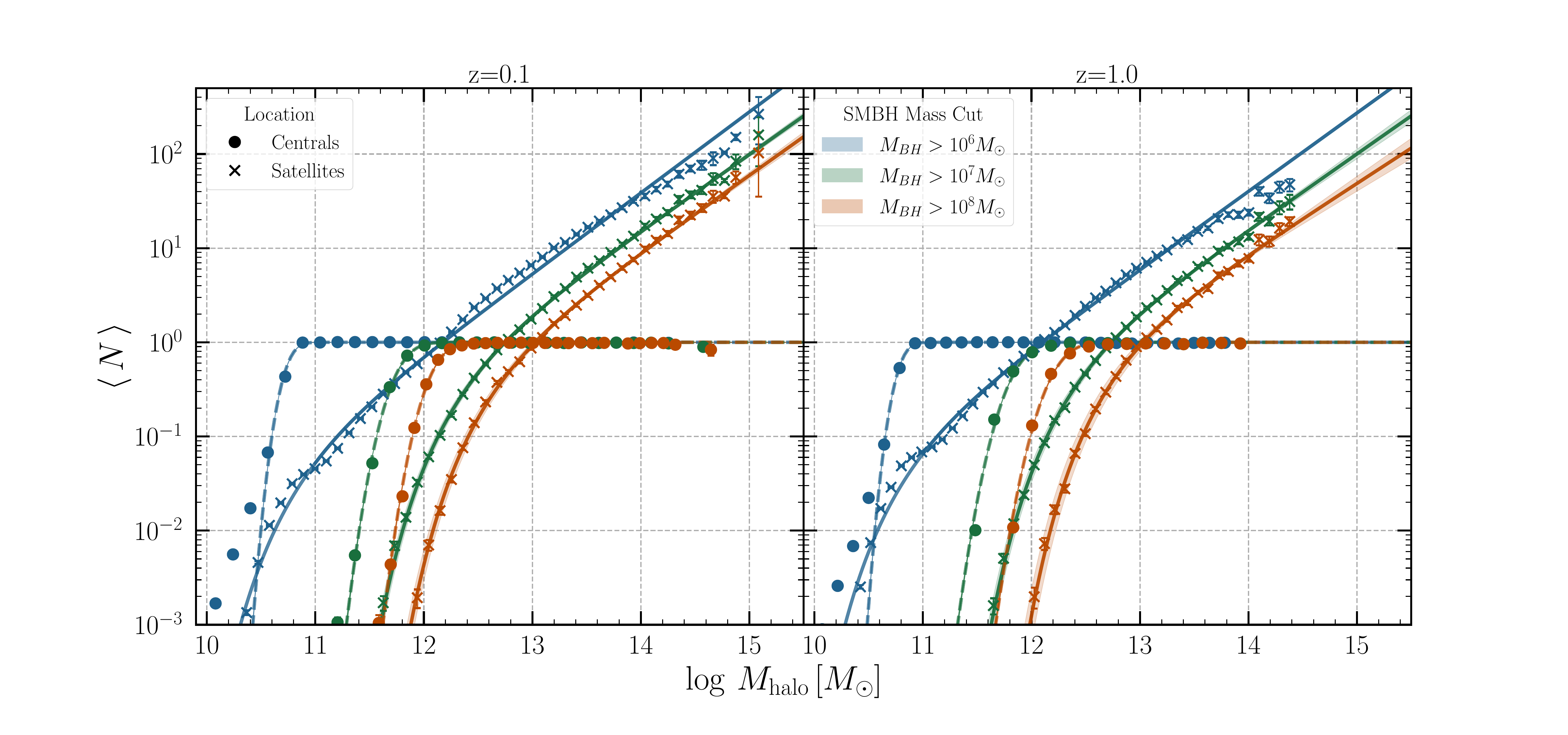}
        \caption{ Mean Occupation Function of mass-selected SMBH in TNG300-1 at redshift, z=0.1 (left) and z=1 (right). The dots (crosses) represent central (satellite) SMBH. The error bars on the data points are $1\sigma$ Poisson error bars. Samples with mass cuts $M_{BH}\,>\,10^{6}\,M_{\odot},\,10^{7}\,M_{\odot},$ and $10^{8}\,M_{\odot}$ are represented in blue, green, and red respectively. The occupation function is well-described by a 5 parameter model as found in previous studies \citep{chatterjeeetal12, degrafetal11b}. The envelopes show the best fit models with 3$\sigma$ errors.}
        \label{fig:Mass_Evolution}
        \end{figure*}

The co-evolution of this galaxy-halo-SMBH system is expected to be regulated by energy and momentum feedback from the central SMBH during their active galacic nuclei (AGN) phase \citep[e.g.,][]{Mountrichas23}. In the last two decades, measurement of the AGN luminosity function \citep[e.g.,][]{Georgakakis15, Fotopoulou16, Ceraj18, Delvecchio20} and the number density of AGNs in the Universe at different redshifts \citep{Soltan82,y&t02,marconietal04} provided insights into the large-scale statistics and accretion histories of AGN. On the other hand, studies of AGN clustering provided a direct way of understanding SMBH evolution along with their host DM halos \citep[e.g.,][]{croometal05, gillietal05, myersetal06, myersetal07a, myersetal07b, shenetal07, shenetal09, Shen13a, Coil07, Coil09, wakeetal08, rossetal09, hickoxetal09, krumpeetal10, krumpeetal12, krumpeetal15, allevatoetal11, hickoxetal11, Cappelluti12, whiteetal12, koutoulidisetal13, mountrichasetal13, eftekharzadehetal15, eftekharzadehetal17}. 

\begin{figure*}[t] 
            \includegraphics[width=\textwidth]{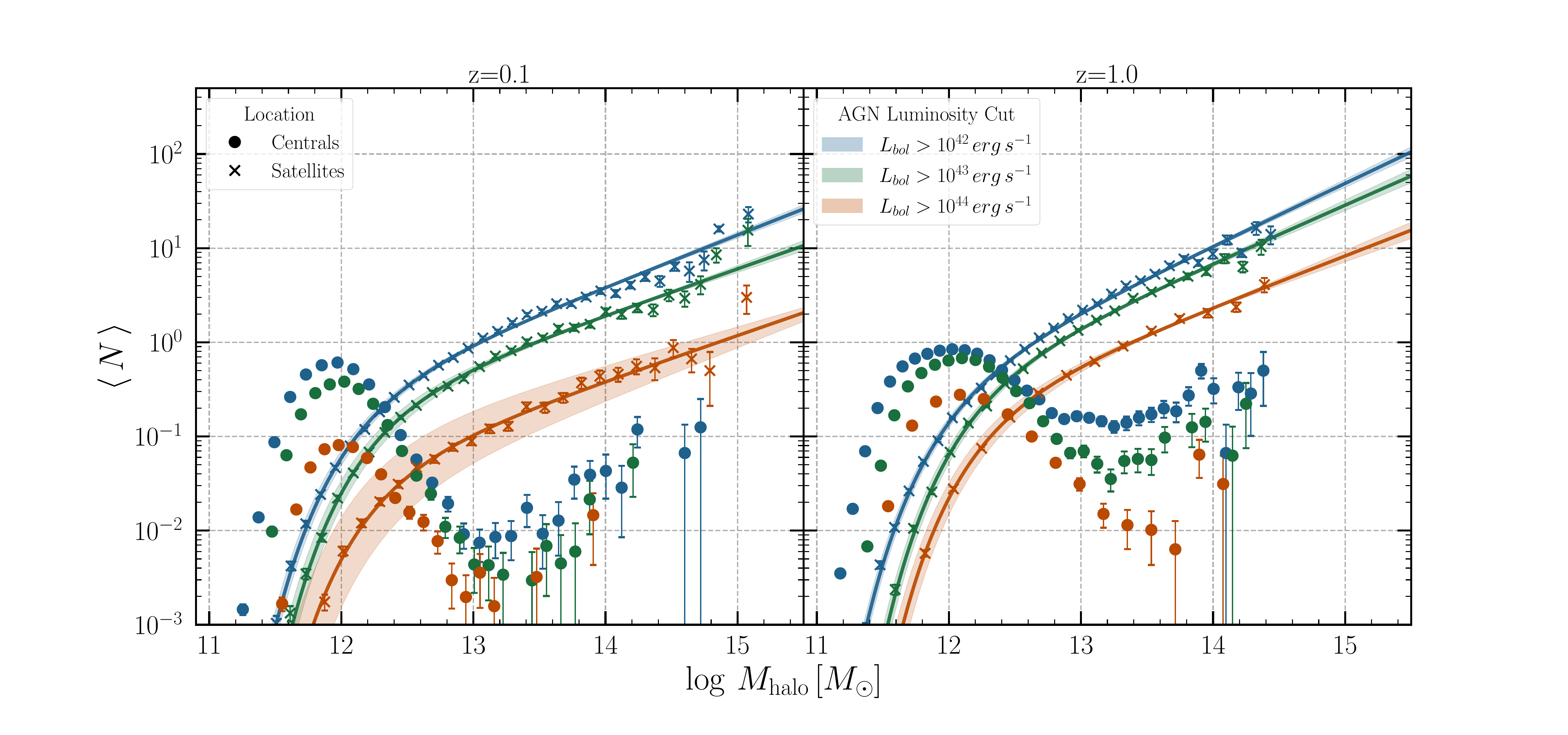}
        \caption{ Mean Occupation Function of luminosity-selected AGNs in TNG300-1 at redshift, z=0.1 (left) and z=1 (right). The dots (crosses) represent central (satellite) AGN. The error bars on the data points are $1\sigma$ Poisson error bars. Samples with luminosity cuts $L_{bol}\,>\,10^{42}\,erg\,s^{-1},\,10^{43}\,erg\,s^{-1}, $ and $ 10^{44}\,erg\,s^{-1}$ are represented in blue, green, and red respectively. The notable departure from the step function (in Fig. \ref{fig:Mass_Evolution}) is evident in the halo mass scales above $10^{12}M_{\odot}$. The envelopes show the best fit $\langle N_{sat} \rangle$ with 3$\sigma$ errors. The best fit parameters are shown in Fig.\ \ref{fig:ParameterVariation.pdf}}
         \label{fig:Lum_Evolution}
        \end{figure*}

  The Halo Occupation Distribution (HOD) framework, which has found widespread application in the interpretation of results from extensive galaxy surveys, links the distribution of galaxies within dark matter halos. The framework is versatile in nature and it provides a simple, yet complete knowledge, of the galaxy-halo connection, since in this approach the galaxy formation physics is singularly linked with the host dark matter halos of galaxies \citep[e.g.,][]{seljak00, b&w02, zhengetal05, z&w07, wakeetal08, shenetal10}. This approach has also been extended to interpret the connection between AGN and the halos they inhabit \citep[e.g.,][]{allevatoetal11, richardsonetal12, k&o12, shenetal12a, richardsonetal13, chatterjeeetal13, dipompeoetal16, chakrabortyetal18, mitraetal18}. One of the biggest challenges in using the HOD framework involves modeling of the HOD of AGN. \citet{chatterjeeetal12} used hydrodynamic simulations \citep{dimatteoetal08} to model the HOD of low-luminosity AGNs. Similar work was carried out by \citet{degrafetal11b} to model the HOD of SMBHs. \citet{bhowmicketal19} utilized the MassiveBlackII \citep{MBII15} simulation with the Conditional Luminosity Function (CLF) framework to extend AGN occupation models for the high-luminosity end, successfully reconciling simulation predictions at low luminosities with observed small-scale quasar clustering.
 
 Using the HOD model of \citet{chatterjeeetal12}, the host dark matter halos of optically bright quasars and X-ray bright AGN were extracted from the two-point correlation function \citep[2PCF;][]{richardsonetal12, richardsonetal13, mitraetal18}. But despite success, these models have extracted widely different physical parameters, such as quasar satellite fractions, while employing the same clustering measurement \citep[e.g.,][]{Shen_2013, graysonetal23, Alametal21,richardsonetal12,Eftekharzadehetal19,Yuanetal24} and hence it is necessary to examine the HOD of AGN spanning the entire luminosity range as well as varied models of black hole accretion and growth. The HOD modeling is potentially sensitive to the subgrid physics (in case of cosmological simulations) or the semi-analytic prescriptions and hence it is important to examine it over a wide range of physical models and a varied set of feedback parameters.

In the last decade, the field of hydrodynamical simulations has made substantial progress in volume, resolution and descriptions of sub-resolution physics\footnote{A quick overview of the current landscape of multiple cosmological hydrodynamical simulations can be found at \url{https://www.tng-project.org/data/landscape/}}.
\citet{DeGraf&Sijacki17} studied the luminosity and redshift dependence of black hole clustering using the Illustris suite of hydrodynamical simulations \citep{Vogelsbergeretal14,geneletal14,Sijaki15}. While modern simulations consistently reproduce multiple galaxy properties at low redshifts, they make divergent predictions for the AGN population \citep{Habouzitetal22}. But, despite limitations, cosmological hydrodynamic simulations are still very useful in studying AGN and their role in galaxy evolution, since they are not tuned to reproduce the properties of AGN, rather they are predictions of the sub-resolution physics built into these simulations in addition to the fact that they provide statistical samples that are useful in cosmological studies. 

In the present work, we use the IllustrisTNG suite of simulations \citep{Springeletal18,Pillepichetal18,Nelsonetal18,Marinaccietal18,Naimanetal18} with the largest simulation TNG300 with a box of side length $\sim$ 300 Mpc to probe the SMBH-halo connection using the HOD formalism by studying its dependence on mass and luminosity. We define two classes of objects, samples of black holes selected by a mass threshold (mass-selected SMBHs) and samples of black holes selected by a luminosity threshold (luminosity-selected AGNs). We compare our results with the SIMBA \citep{dave19} suite of simulations, which has side length of $\sim$147 Mpc. We investigate the effect of different AGN feedback modes used in SIMBA, on the HOD, using their smaller simulation runs of $\sim$75 Mpc box size. We directly calculate the Mean Occupation Function (MOF) of mass-selected SMBH and luminosity-selected AGN samples in multiple redshift slices of these simulations to probe the redshift evolution. 

We find that the luminsoity-selected AGN population is suppressed at a characteristic halo mass scale ($\sim 10^{13}M_{\odot}$) affecting the nature of its occupation distribution function from what has been found before \citep[e.g.,][]{zhengetal05, z&w07, chatterjeeetal12}. We  further investigate the effect of this suppression in clustering measurements. Our results show that, without adequate modeling of the physical effects in the HOD, we are likely to extract a significantly different distribution of host halo masses of AGN, which will impact our understanding of their co-evolution paradigm. The present draft is organized as follows. In \S \ref{sec:simulation} we describe the simulations, our sample selection, and the standard 5 parameter model. In \S \ref{Results} we discuss the analysis techniques used and present our results. In \S \ref{Discussions} we discuss and summarize our findings. Throughout our work we use a $\Lambda$CDM cosmology with parameters from \citet{planckcollaborationetal13XI}.

\section{Simulations}		\label{sec:simulation}
We now briefly describe the two simulations used in this work, namely, IllustrisTNG and SIMBA in the next sections. 
\subsection{IllustrisTNG}

The Next Generation Illustris simulations (IllustrisTNG) are a suite of hydro-dynamical simulations carried on large cosmological volumes. IllustrisTNG have been carried out in three different box sizes. In this work we use TNG300 and TNG100 which have periodic box sizes of L $\sim 205 h^{-1}Mpc$ and L $\sim 75 h^{-1}Mpc$ respectively. Initial conditions are consistent with \citet{Planck16}. IllustrisTNG uses the moving-mesh cosmological simulation code AREPO \citep{Springel10,Weinberger20}. Gravity is solved using a Tree-Particle-Mesh (Tree-PM) method and for magneto-hydrodynamics a quasi-Lagrangian method has been used. Gravity is treated completely in the Newtonian framework with periodic boundary conditions. 

The galaxy physics directly follows the framework used in Illustris simulations. IllustrisTNG however has updated descriptions for growth and feedback of SMBH, galactic winds, stellar evolution and chemical enrichment, among other important physics updates. For more details, we refer the reader to the IllustrisTNG repository \footnote{\url{https://www.tng-project.org/}} and references therein. An on the fly friends-of-friends (FoF) halo finder identifies the halos. Seed black holes of mass 8 $\times 10^{5} h^{-1} M_{\odot}$ are placed at the centers of halos whenever the halo crosses a threshold mass of $5 \times 10^{10} h^{-1} M_{\odot}$, and does not yet have a black hole. The black hole then dynamically grow by accretion and feeds back energy into the surrounding medium. For details about the accretion and feedback models used in IllustrisTNG refer to \citet{Springeletal18,Pillepichetal18,Nelsonetal18,Marinaccietal18,Naimanetal18,weinberger17a,weinbergeretal18}.

\begin{figure*}[t] 
            \includegraphics[width=\textwidth]{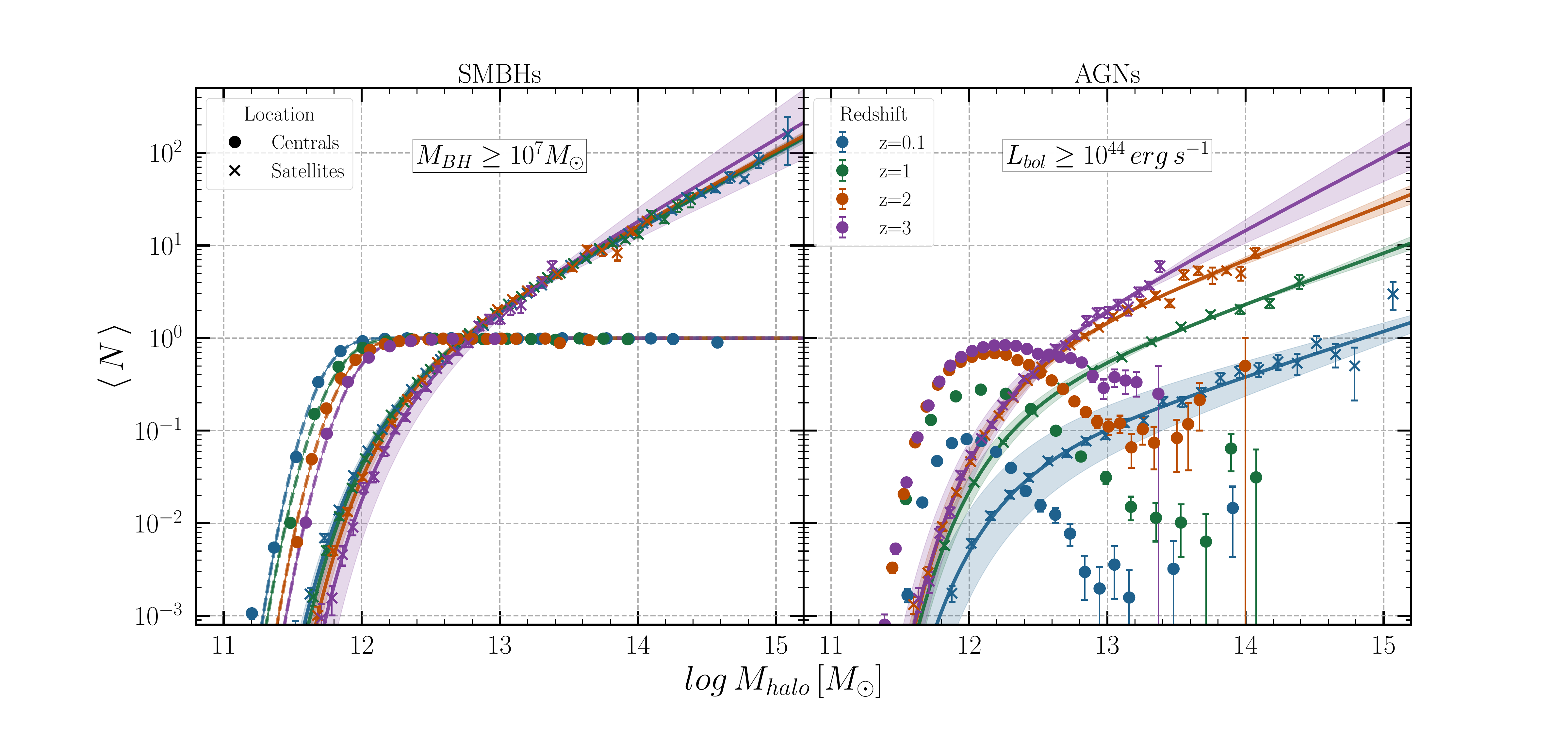}
        \caption{ \textbf{Left :} Redshift evolution of the Mean Occupation Function of mass-selected SMBHs with $M_{BH}>10^{7}\,M_{\odot}$ in TNG300-1. The dots (crosses) represent the central (satellite) SMBH. The error bars of data points are $1\sigma$ Poisson error bars. MOF at $z=0.1, 1, 2, 3$ are shown in blue, red, green, and purple respectively. \textbf{ Right :} Redshift evolution of the Mean Occupation Function of luminosity-selected AGN with $L_{bol}\,>\,10^{44}\,erg\,s^{-1}$ in TNG300-1. The dots (crosses) represent the central (satellite) AGN. The error bars of data points are $1\sigma$ Poisson error bars. MOF at z=0.1,1,2,3 are shown in blue, red, green, purple respectively. The decrease in AGN fraction at a characteristic mass scale of $10^{13}M_{\odot}$ is evident at all redshifts although the effect is much stronger at lower redshifts. The envelopes show the best fit with 3$\sigma$ errors.}
        \label{fig:Redshift_Evolution_SMBH}
        \end{figure*}

\begin{figure}[t] 
            \includegraphics[width=\columnwidth]{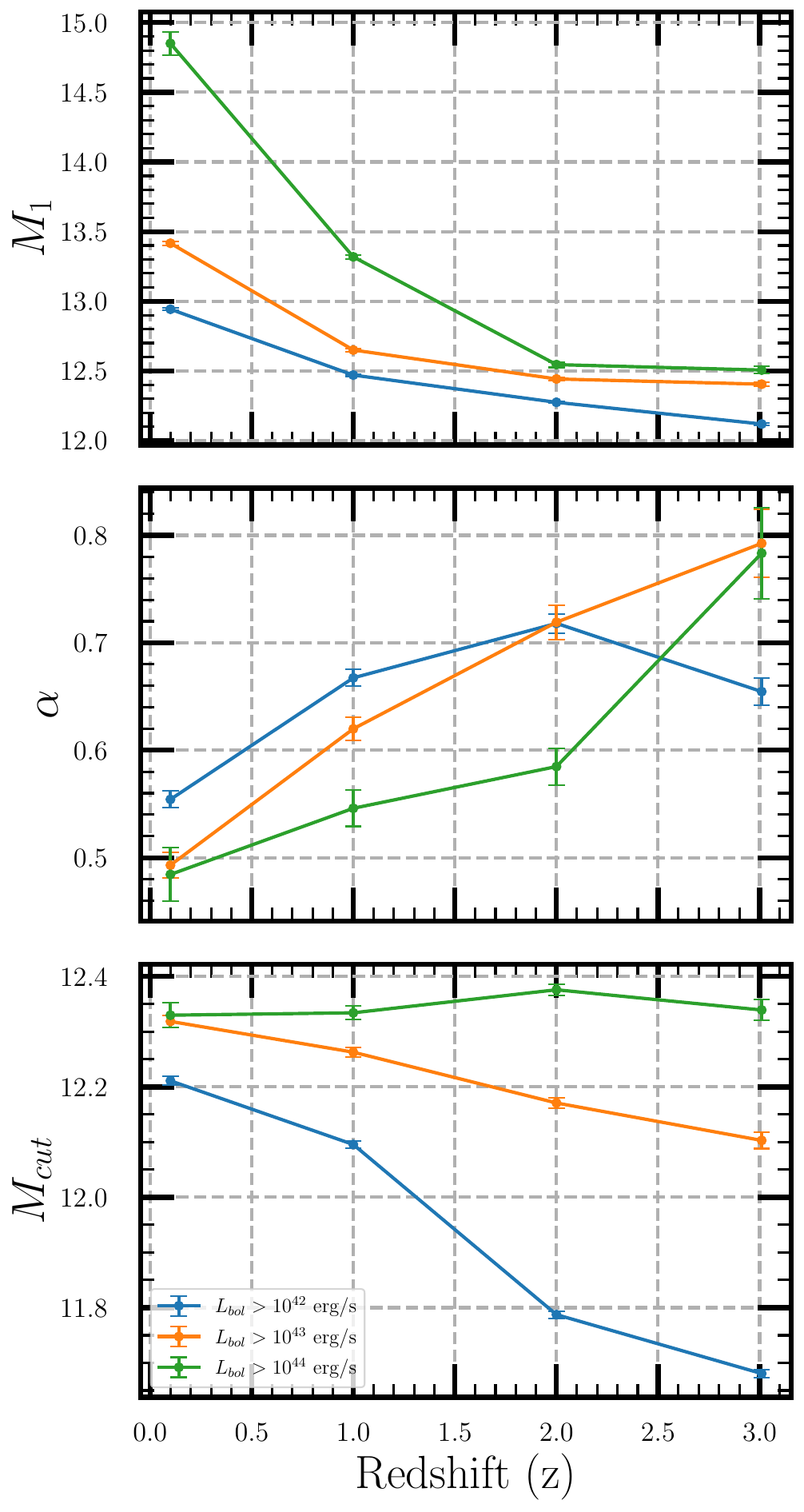}
        \caption{Evolution of the best fit parameters of Eq.\ref{eqn:satellite} as a function of redshift(z). Parameters for three different luminosity cuts of AGNs, $L_{bol}\,>\,10^{42}\,erg\,s^{-1},\,10^{43}\,erg\,s^{-1}, $ and $ 10^{44}\,erg\,s^{-1}$ are represented in blue, orange, and, green respectively. \textbf{Top :} Variation of $M_1$ with z. \textbf{Middle :} Variation of $\alpha$ with z. \textbf{Bottom :} Variation of $M_{cut}$ with z.} 
        \label{fig:ParameterVariation.pdf}
        \end{figure}
\subsection{SIMBA}
The SIMBA simulation \citep{dave19} is an update over the MUFASA simulations \citep{dave2016mufasa}. The initial conditions are generated with the parameters consistent with \citet{planckcollaborationetal13XI}. The simulation begins at z= 249 and evolves until z = 0. SIMBA runs on a modified version of the GIZMO \citep{hopkins2015new} cosmological gravity plus hydrodynamics solver. GIZMO \citep{GIZMO15} is based on GADGET-3, but includes significant updates, most notably the meshless finite mass (MFM) hydrodynamics solver, which improves the accuracy in capturing fluid mixing and shocks compared to the traditional SPH method used in GADGET-3 \citep{Springel05Gadget}. SIMBA uses GIZMO in its MFM version.  DM and gas particles evolve simultaneously under gravity and pressure forces, and shocks are handled via a Riemann solver with no artificial viscosity \citep{dave2016mufasa,dave19}.
 
An on-the-fly FoF algorithm is used to seed black holes in galaxies dynamically. If a galaxy reaches a stellar mass, $M_{*} \gtrsim 10^{9.5} M_{\odot} $, the star particle closest to the center of mass of the galaxy is converted into a black hole particle of mass $M_{seed} = 10^{4} h^{-1}M_{\odot} $. The black holes are then allowed to grow using two modes of accretion: Torque limited accretion of cold gas and Bondi accretion of hot gas. The energy from accretion is used to drive feedback, which quenches galaxies. For feedback, a kinetic sub-grid model is incorporated along with an X-ray energy feedback. At high Eddington ratio ($f_{edd}$) mode outflows, radiative AGN winds provide feedback and as $f_{edd}$ starts to drop ($f_{edd}<0.2$), the outflow transitions into a jet mode of feedback. The X-ray mode of feedback incorporates the heating of gas due to energy input from X-ray off the accretion disc.

\subsection{Sample Selection} \label{samples}
We use group catalogs provided by the TNG collaboration containing a halo catalog and a SUBFIND subhalo (galaxy) catalog to construct the host DM halo and the corresponding mass-selected SMBH and luminosity-selected AGN samples. For the mass estimate of halos, we use $M_{200}$, which is defined as the mass enclosed by the region where the density drops to 200 times the critical density of the Universe. SUBFIND subhalo catalog uses the SUBFIND algorithm \citep{Springel05Gadget} to identify galaxies in the simulation. 

For SIMBA, we use the CAESAR catalog, which is an extension of the Python-based simulation analysis toolkit \textsc{yt} \citep{Turketal11}. CAESAR utilizes a 6-degree FoF algorithm to identify halos and galaxies in the simulation, and it also cross-matches the two catalogs to determine the parent halos of each galaxy. For halo mass, we again use the $M_{200}$ estimate from the CAESAR catalog. Similarly, for the mass-selected SMBH and luminosity-selected AGN sample, we use galaxies from the CAESAR catalog.

In this work, for both TNG and SIMBA, we consider halos above a threshold mass of $10^{10}\, M_{\odot}$ to avoid resolution effects. To avoid artifacts that might be caused because of black-hole seeding, we consider only the black-holes above a threshold mass of, $M_{BH}>\,10^{6}\,M_{\odot}$ assuming that every black-hole in our sample has had enough time to evolve naturally. In our analysis we use snapshots from both simulations at redshifts, $z=3, 2, 1, 0.1$. The bolometric luminosity of the BHs in both simulations are calculated using the prescriptions of \citet{Habouzitetal22}, where we distinguish between radiatively efficient AGN with $f_{edd} > 0.1$ and bolometric luminosity given as 
\begin{equation}
    L_{bol} = \frac{\epsilon_{r}}{1-\epsilon_{r}}\dot{M}_{BH}c^{2}.
\end{equation}
For radiatively inefficient AGN with $f_{edd} \leq 0.1$ the bolometric luminosity is calculated as
\begin{equation}
    L_{bol} = (10f_{edd})\epsilon_{r}\dot{M}_{BH}c^{2}
\end{equation}
We chose the radiative efficiency parameters ($\epsilon_{r}$) from IllustrisTNG and SIMBA, which are used to calibrate the $M_{BH}-M_{*}$ scaling relations. We take $\epsilon_{r} = 0.2$ for IllustrisTNG and $\epsilon_{r}=0.1$ for SIMBA. The Eddington ratio ($f_{edd}$) is computed from the simulations, using the instantaneous value of the accretion rate and the mass of the black hole. 

In addition to our AGN sample, we construct a galaxy sample from the simulation, consisting star-forming and quenched galaxies. To  construct the sample, we use the criteria based on \citet{donnarietal19} where we divide the entire galaxy sample based on a specific star formation rate cut (sSFR). We define our star-forming and quenched galaxies to have $\rm log\,\,sSFR(yr^{-1}) \geq -11$ and $\rm log\,\,sSFR(yr^{-1}) \leq -11$ respectively. We impart a halo mass cut of $10^{10}M_{\odot}$ in constructing our galaxy sample similar to what has been done for the AGN sample.

\subsection{Central and Satellite AGN}  \label{MOF}The first moment of the HOD, the mean of the distribution, is assumed to be a function of halo mass $\langle N (M_{halo}) \rangle$ only, being independent of any other secondary properties \citep{b&w02, berlindetal03}. The MOF of galaxies and AGN is characterized by physically motivated models that describe the distribution of these objects in halos as the sum of central and satellite contributions Eq. \ref{eqn:Total} \citep{zhengetal05, chatterjeeetal12}. In TNG, we assume the most massive galaxy inside a given halo to be the central galaxy and the SMBH it hosts to be the central SMBH. In SIMBA, we use the galaxy with the highest stellar mass as the central galaxy and its SMBH as the central SMBH. By this definition, the number of central SMBH inside a given halo can either be zero or one.

  Traditionally the Mean Occupation Function (MOF) of central ($\left \langle N_{cen}(M) \right \rangle$) SMBH is modeled as a softened step function (Eq.\ \ref{eqn:centralBH}). $M_{min}$ gives us the halo mass at which the average number of central SMBH for the given halo mass is 0.5; $\sigma_{\log M}$ characterizes the typical transition width of the softened step function. The MOF of satellite SMBH is modeled as a power law with an exponential roll-off (Eq. \ref{eqn:satellite}) at smaller halo mass scales. The parameter $M_{1}$ characterizes the typical host halo mass required to host a satellite SMBH; $\alpha$ is the power law index; $M_{cut}$ is used to model the rolling of the power law. We use this standard 5-parameter model to test for its validity in describing AGN HOD (see sections \ref{sec:Mass and Luminosity Evolution} and  \ref{sec:clustering}).
\begin{equation} \label{eqn:Total}
\left \langle N(M) \right \rangle = \left \langle N_{cen}(M) \right \rangle + \left \langle N_{sat}(M) \right \rangle
\end{equation}
\begin{equation}	\label{eqn:centralBH}
\left \langle N_{cen} \right \rangle = \frac{1}{2} \left ( 1 + erf \bigg ( \frac{\log M-\log M_{min}}{\sigma_{\log M}} \bigg ) \right ) \\
\end{equation}
\begin{equation}	\label{eqn:satellite}
\left \langle N_{sat}(M) \right \rangle =\ \Bigg(\frac{M}{M_1}\Bigg)^{\alpha}\exp{\Bigg(-\frac{M_{cut}}{M}\Bigg)}
\end{equation} 

\begin{figure}[t] 
            \includegraphics[width=\columnwidth]{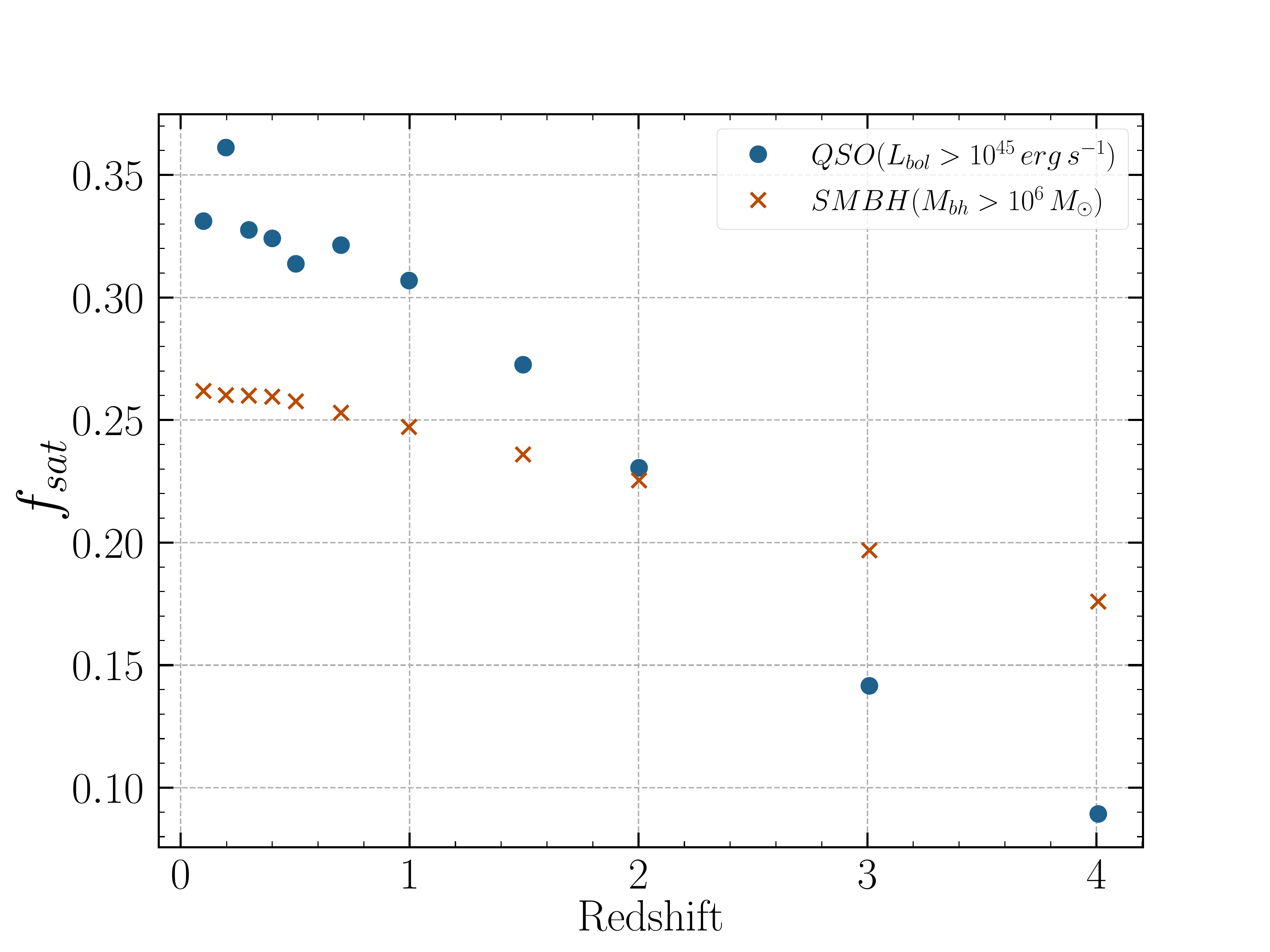}
        \caption{Global satellite fraction of luminosity-selected AGNs (Quasars : $L_{bol} > 10^{45}$ ergs s$^{-1}$) and mass-selected SMBHs (SMBHs : $M_{bh} > 10^{6}$ $M_{\odot}$) as a function of redshift in TNG300-1. The satellite fraction of our quasar sample increases as we go to lower redshifts while it saturates below $z=1.0$ for the SMBH sample. Our findings support the results of \citet{Alametal21} who show using clustering studies that $20-40\%$ of QSO sin the intermediate redshift ranges are satellites. See \S 4 for discussions. }
        \label{fig:fsat}
        \end{figure}

\section{Results}  \label{Results}
We compute the HOD of SMBH selected in our simulations and investigate its variations over black hole parameter spaces, namely mass, luminosity, redshift and feedback modes. We present our results as follows. 

\subsection{Mass Luminosity and Redshift Evolution} \label{sec:Mass and Luminosity Evolution}

The MOF of mass-selected SMBHs in halos is expected to evolve with black hole mass. To study the mass evolution of the MOF, we construct three samples of SMBHs by taking mass cuts of $10^{6}$ $M_{\odot}$, $10^{7}$ $M_{\odot}$ and $10^{8}$ $M_{\odot}$ at different redshifts. In Fig. \ref{fig:Mass_Evolution} we show $\langle N_{cen} \rangle$ and $ \langle N_{sat} \rangle$ at redshifts z=1 and z=0.1 for the TNG300-1 simulation. We observe that host halo mass scales increase with the threshold mass of black holes, for both central and satellite populations, representing the halo mass - black hole mass correlation \citep{ferrarese02, volonterietal11, baesetal03, tremaineetal02}.  The MOF represents a softened step function for the central SMBH, as has been described in Eq.\ \ref{eqn:centralBH}. The satellite population follows a power-law behavior (Eq.\ \ref{eqn:satellite}). 

To study the luminosity evolution of the AGN MOF, we take three samples of our luminosity-selected AGNs with cuts of $L_{bol}$  $10^{42}$ erg s$^{-1}$, $10^{43}$ erg s$^{-1}$ and $10^{44}$ erg s$^{-1}$. In Fig. \ref{fig:Lum_Evolution} we plot $\langle N_{cen} \rangle$ and $\langle N_{sat} \rangle$ for AGN at redshift z=1 and z=0.1. The behavior of the central occupation is significantly different from our mass-selected SMBH samples in Fig. \ref{fig:Mass_Evolution}. At both redshifts, $\langle N_{cen} \rangle$ goes to a minimum at a typical host halo mass scale ($\sim 10^{12.5-13}M_{\odot}$). The central occupation ($\langle N_{cen} \rangle$) could not be fit using the usual step function contrary to previous studies. The satellite population follows a power-law like behavior.

\begin{figure*}[t] 
  \includegraphics[width=18cm]{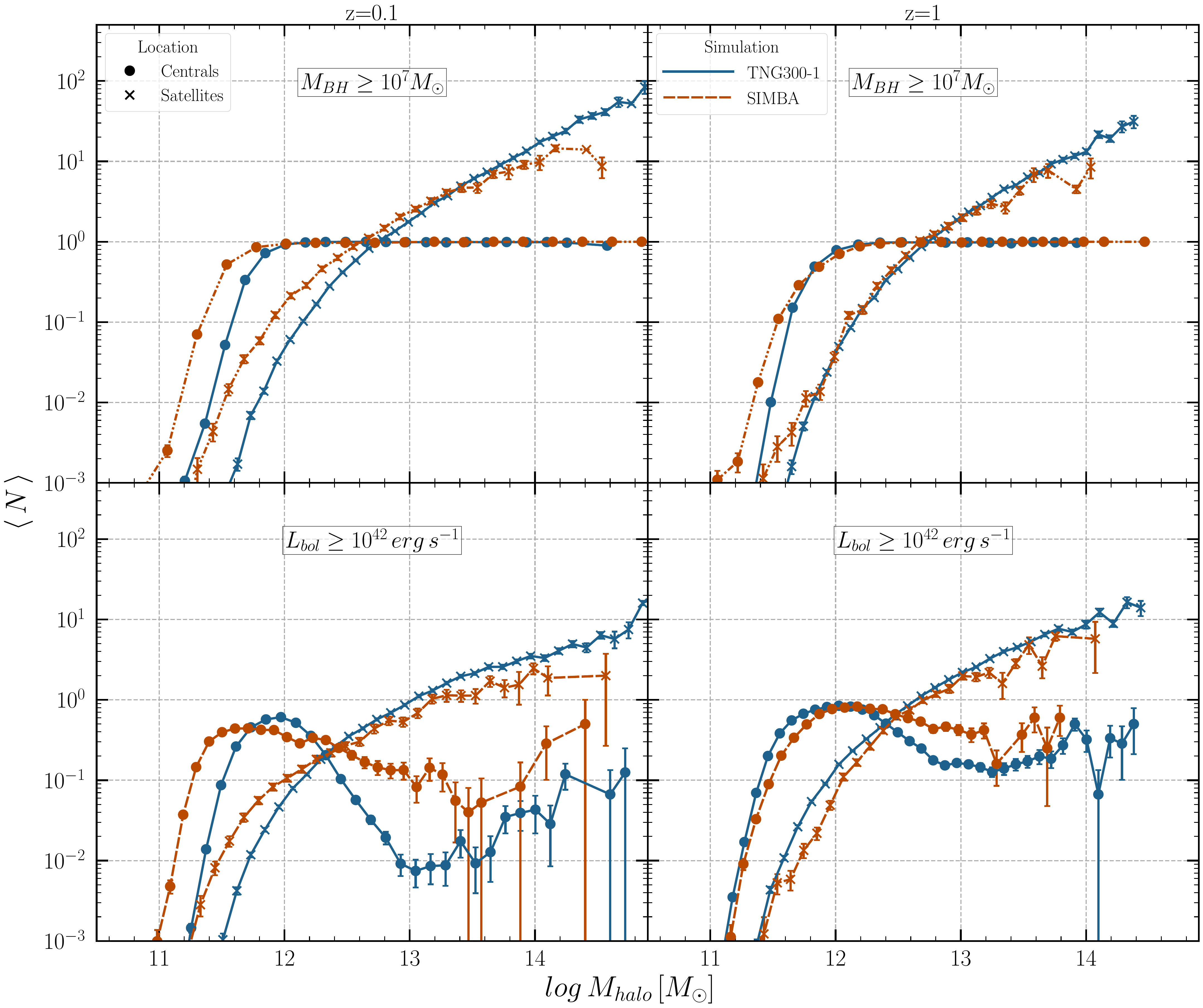}
  \caption{\textbf{Top} :  Comparison between Mean Occupation Function of mass-selected SMBH in TNG300-1 and SIMBA at redshifts z=0.1 (left) and z=1 (right). Dots (crosses) represent central (satellite) SMBH with a mass cut of $M_{BH}>\,10^{7}\,M_{\odot}$. \textbf{Bottom} : Comparison between Mean Occupation Function of luminosity-selected AGNs in TNG300-1 and SIMBA at redshifts z=0.1 (left) and z=1 (right). Dots (crosses) represent central (satellite) AGN with a luminosity cut of $L_{bol}$  $>10^{42}$ erg s$^{-1}$. The error bars on data points are $1\sigma$ Poisson error bars. AGNs in TNG300-1 and SIMBA are represented in blue and red respectively. } 
  \label{fig:TNG_v_SIMBA}
\end{figure*}

To study how the mass-selected SMBHs and luminosity-selected AGNs are distributed in halos at different redshifts, we compute the MOF for different luminosity and mass cuts, at different redshift slices. In Fig. \ref{fig:Redshift_Evolution_SMBH}  we plot the MOF of our mass-selected SMBH and luminosity-selected AGN samples at redshift slices of z= 0.1, 1, 2 and 3. The MOF of mass-selected SMBHs remains unchanged across redshifts. (left panel of \ref{fig:Redshift_Evolution_SMBH}). For our luminosity-selected AGN sample (right panel of \ref{fig:Redshift_Evolution_SMBH}) we note that $\langle N_{cen} \rangle$ substantially evolves with redshift and consistently exhibits a dip at a halo mass scale of $\sim 10^{13}M_{\odot}$. As we go to higher redshifts the MOF flattens to attend the step-function like behavior. The satellite occupation fraction ($\langle N_{sat} \rangle$) evolves relatively weakly at higher redshifts with the strongest evolution being observed between redshift z$=1$ and z$=0.1$. No such redshift dependence is observed for the mass-selected SMBH sample (left panel of Fig. \ref{fig:Redshift_Evolution_SMBH}). 
In Fig \ref{fig:ParameterVariation.pdf} we plot the best fit parameters for $\langle N_{sat} \rangle$ as a function of redshift for our luminosity-selected AGNs. We find that the parameter \( M_1 \) exhibits a strong redshift dependence for the most luminous AGNs. At low redshifts, the typical host halo mass required to host satellite AGNs with \( L_{\mathrm{bol}} \geq 10^{44}\, \mathrm{erg/s} \) is approximately two orders of magnitude higher than that for AGNs with \( L_{\mathrm{bol}} \geq 10^{42}\, \mathrm{erg/s} \). In contrast, at high redshifts, the typical host halo masses are comparable across all luminosity thresholds. This suggests that high-luminosity satellite AGNs require rarer, more massive environments. However, at high redshifts, this luminosity dependence weakens, indicating that luminous satellite AGNs can reside in a broader range of halo masses—possibly due to higher gas fractions and more frequent galaxy interactions. The parameter \( M_{\mathrm{cut}} \) remains roughly constant for the most luminous AGNs but shows variation for the \( L_{\mathrm{bol}} \geq 10^{42}\, \mathrm{erg/s} \) sample, indicating that the cutoff in satellite AGN occupation occurs at somewhat higher halo masses in the low-redshift universe compared to high redshift, consistent with an environment-dependent quenching scenario. The power-law slope \( \alpha \) generally increases with redshift for all three luminosity samples, consistent with the scenario of secondary effects (e.g., mergers) contributing to the satellite population. The distribution of the satellites are close to a Poisson distribution and the results for a representative set are shown in Fig.\ \ref{fig:poisson_fitted_plot.pdf}  (Appendix).

\begin{figure*}[t] 
  \includegraphics[width=17cm]{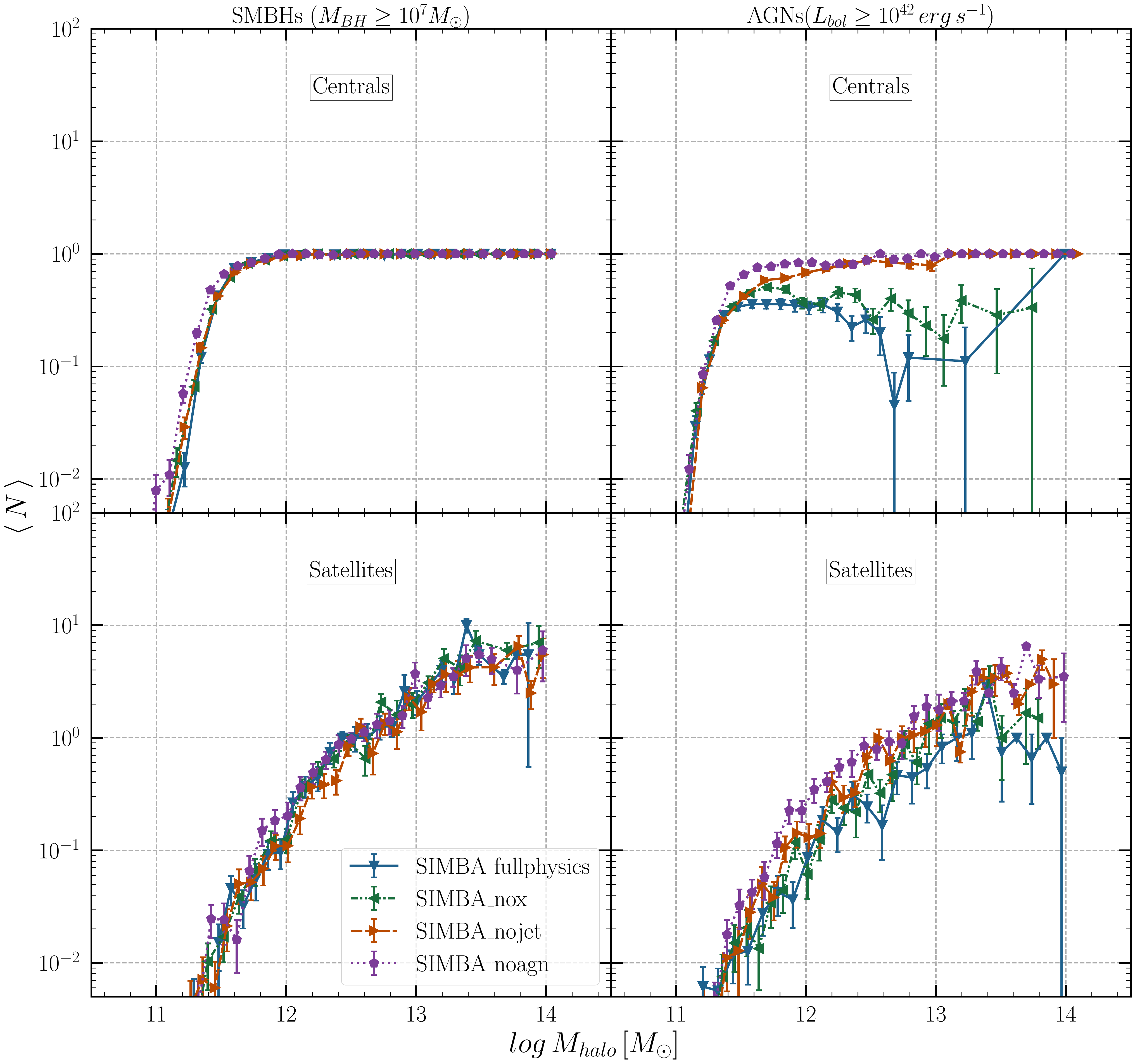}\\
  \caption{\textbf{Left} : Mean Occupation Function of central (top) and satellite (bottom) SMBH, with a mass cut of $M_{BH}>\,10^{7}\,M_{\odot}$ in different SIMBA simulation runs at z=0.1. \textbf{Right} : Mean Occupation Function of central (top) and satellite (bottom) AGN, with a luminosity cut of $L_{bol}$  $>10^{42}$ erg s$^{-1}$ in different SIMBA simulation runs at z=0.1. While the different feedback modes have almost no effect on the occupation function of black holes, it significantly affects the occupation function of central AGN in halos. The departure from the step-function is evident when the jet mode of feedback (green and the blue triangles in the top right panel) is introduced in SIMBA. }
  \label{fig:SIMBA_Feedback}
\end{figure*}

Complementary to Fig. \ref{fig:ParameterVariation.pdf}, in Fig.~\ref{fig:fsat}, we directly compare the satellite fraction of the quasar sample (defined as $L_{bol} >10^{45}$ erg s$^{-1}$) to that of a mass-selected control sample of SMBHs ($M_{bh} >10^{6} {\rm M_{\odot}}$). While both quantify the fraction of black holes residing in satellite galaxies, they are defined over fundamentally different selections: one based on luminosity, the other on mass. We find that the satellite fraction for quasars increases approximately linearly from $z=4$ to $z=0.1$, whereas the satellite fraction for the mass-selected SMBH sample flattens below $z=1.0$. This difference highlights a clear dependence on luminosity and redshift in satellite occupation. The results are in agreement with \citet{Alametal21} who report $f_{sat}$ to be $\sim$20 to $\sim$40 percent for quasars at intermediate redshifts. For more discussion, see \S 4.

\subsection{Evolution with Feedback Modes}
 We use TNG300-1 to compare the results with SIMBA (Fig.\ref{fig:TNG_v_SIMBA}). We note that both SIMBA and TNG produce similar results for MOF for the mass-selected SMBH samples (top panel) except for the low mass end of the satellite population at z=0.1. The result is slightly different for the luminosity-selected AGN samples when divided in central and satellites (lower panels). We note that the total occupation fraction converges quite well between SIMBA and TNG-300-1. 
 
 It is interesting to note that both SIMBA and TNG show a dip in the central occupation at the halo mass scale of $\sim 10^{13}M_{\odot}$, but slight disagreement in the satellite population at both redshifts is evident from the results. However, $\langle N_{sat} \rangle$ in both cases, is well explained by Eq.\ \ref{eqn:satellite}. In TNG, black hole accretion is modeled solely through a Bondi prescription, assuming spherical accretion from the surrounding gas. In contrast, SIMBA distinguishes between cold and hot gas, using a torque-limited accretion model (GTDA) for cold, rotationally supported gas and Bondi accretion for hot gas, leading to more efficient black hole fueling in dense, star-forming regions. 

As discussed before, SIMBA has different modes of feedback and the flagship run is the one which includes all the modes. We now compute the MOF for different modes of feedback in SIMBA (Fig.\ref{fig:SIMBA_Feedback}). SIMBA uses three modes of AGN feedback. Black holes with a high Eddington ratio ($f_{edd}$) generate kinetic outflows (AGN winds) into the inter stellar medium according to Eq.\ \ref{eqn:wind}. 

\begin{equation}
\label{eqn:wind}
   v_{w,EL} = 500 + 500 (log M_{BH}-6)/3\,km s^{-1} 
\end{equation}
At lower Eddington ratios, $f_{edd}\,\lesssim0.02$, the outflow transitions to a jet mode, with outflow velocity increasing with the Eddington rate given by, Eq.\ \ref{eqn:jet}. 
\begin{equation}\label{eqn:jet}
    v_{w,jet} = v_{w,EL} + 7000(0.2/f_{edd})\,km s^{-1}
\end{equation}
To model the feedback from the photons that reflect from the accretion disc (X-ray mode), SIMBA inputs energy into the surrounding using a radiative efficiency of 0.1. The X-ray mode turns on only when the jet mode has turned on. We use the smaller SIMBA boxes with different feedback modes to compare the effects of different modes of AGN feedback on the MOF. The results are shown in Fig.\ref{fig:SIMBA_Feedback}. SIMBA\_fullphysics is a smaller 75 Mpc box with all AGN modes of feedback turned on. The X-ray mode is switched off in SIMBA\_nox, while jet mode and AGN winds remain on. SIMBA\_nojet has neither X-ray nor jet mode, AGN winds are the only mechanism for feedback. In SIMBA\_noagn there is no AGN feedback. Stellar feedback is present in all the simulation boxes. In Fig.\ref{fig:SIMBA_Feedback} we plot the MOF calculated at z=0.1 for all the simulation runs. We see that the feedback modes do not have any effect on the distribution of mass-selected SMBHs (left panels), while the MOF of luminosity-selected AGN samples vary with varying models of feedback. We note that once the jet mode is turned on, both the central and the satellite occupations get suppressed. The results stay similar for the $z=1.0$ case.

\begin{figure*}[t] 
  \includegraphics[width=18cm]{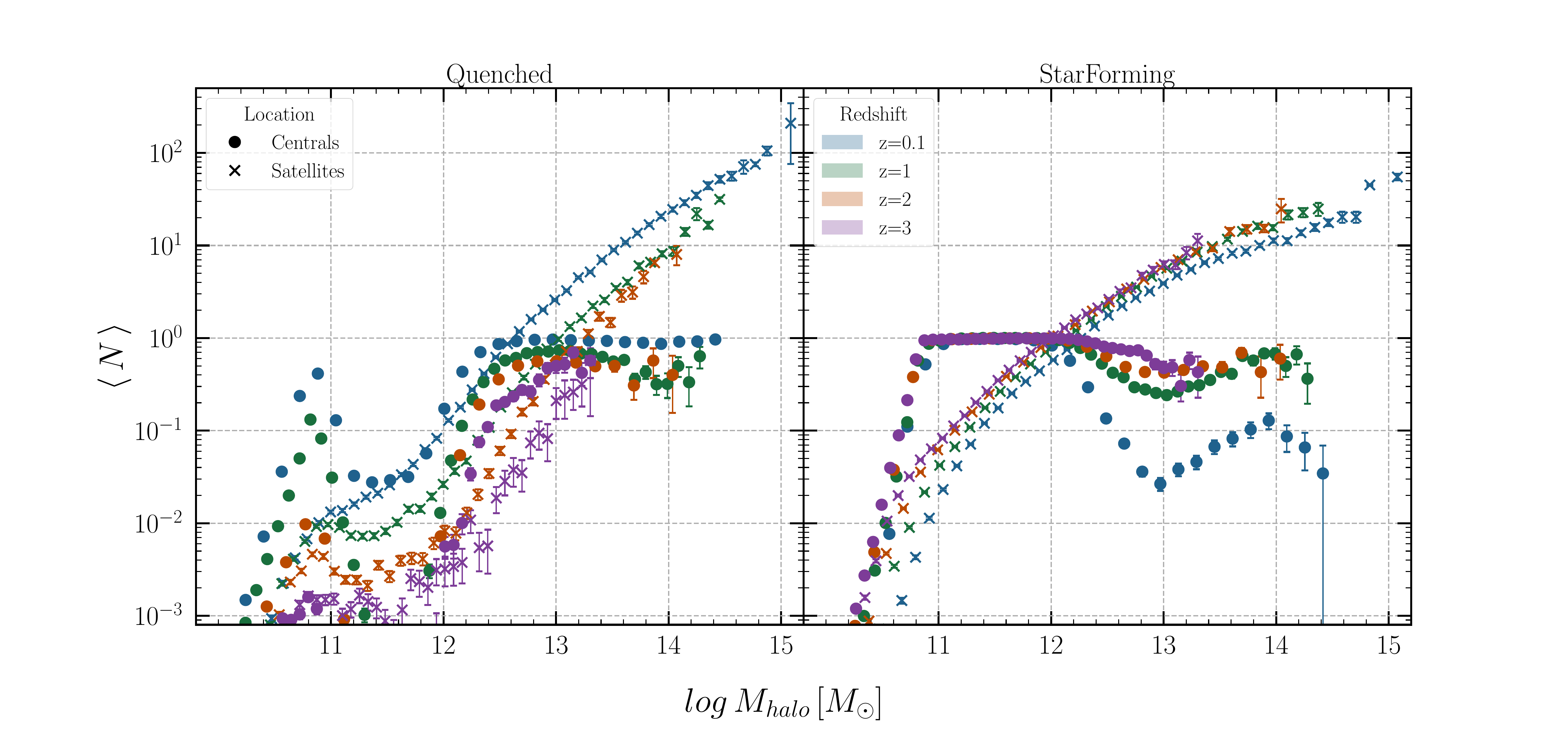}
  \caption{\textbf{Left :} Redshift evolution of the Mean Occupation Function of Quenched Galaxies with $log\,\,sSFR(yr^{-1}) < -11$ in TNG300-1. The dots (crosses) represent the central (satellite) SMBH. The error bars of data points are $1\sigma$ Poisson error bars. MOF at $z=0.1, 1, 2, 3$ are shown in blue, red, green, and purple respectively. \textbf{ Right :} Redshift evolution of the Mean Occupation Function of Star Forming Galaxies with $log\,\,sSFR(yr^{-1}) \geq -11$ in TNG300-1. The dots (crosses) represent the central (satellite) AGN. The error bars of data points are $1\sigma$ Poisson error bars. MOF at z=0.1,1,2,3 are shown in blue, red, green, purple respectively. The decrease in fraction of star forming galaxies, similar to AGN fraction, at a characteristic mass scale of $10^{13}M_{\odot}$ is evident at all redshifts although the effect is much stronger at lower redshifts.}
  \label{fig:Star_Formation_HOD}
  \end{figure*}

\subsection{Star Forming Galaxies}
As discussed, the effect of feedback is evident on the occupation statistics of AGN/quasars, and we consistently see a dip in the MOF of luminosity-selected AGN at a characteristic halo mass scale. To further investigate the effect, we now compute the MOF of star-forming galaxies in the simulation.

We extracted the MOF for our two samples of quenched and star-forming galaxies as per the criterion defined in \S \ref{samples}. In Fig. \ref{fig:Star_Formation_HOD} we show the HOD of quenched (left) and star-forming galaxies (right) at different redshifts. We find that the MOF of central star-forming galaxies show a trend similar to that of central AGNs where we see a dip in the MOF at a halo mass scale of $10^{13}M_{\odot}$. The MOF is significantly suppressed at lower redshift for the star-forming population. No such effect of suppression of the occupation fraction at characteristic halo masses or at low redshifts is seen in the case of quenched galaxies. Simultaneous quenching of AGN and star-formation reinforces the idea that AGN feedback (or similar modeling parameters) play a key role in regulating both star formation and black hole growth in central galaxies of similar halo environments. It is thus likely that similar modeling parameters in the simulation are simultaneously responsible for shutting down star-formation and suppressing AGN accretion at a typical host halo scale. In our follow-up work (Chowdhary et al. 2025 in preparation) we propose to characterize the host galaxies of our AGNs based on specific star-formation and investigate the AGN-host galaxy connection.

\begin{figure*}[t] 
  \includegraphics[width=18cm]{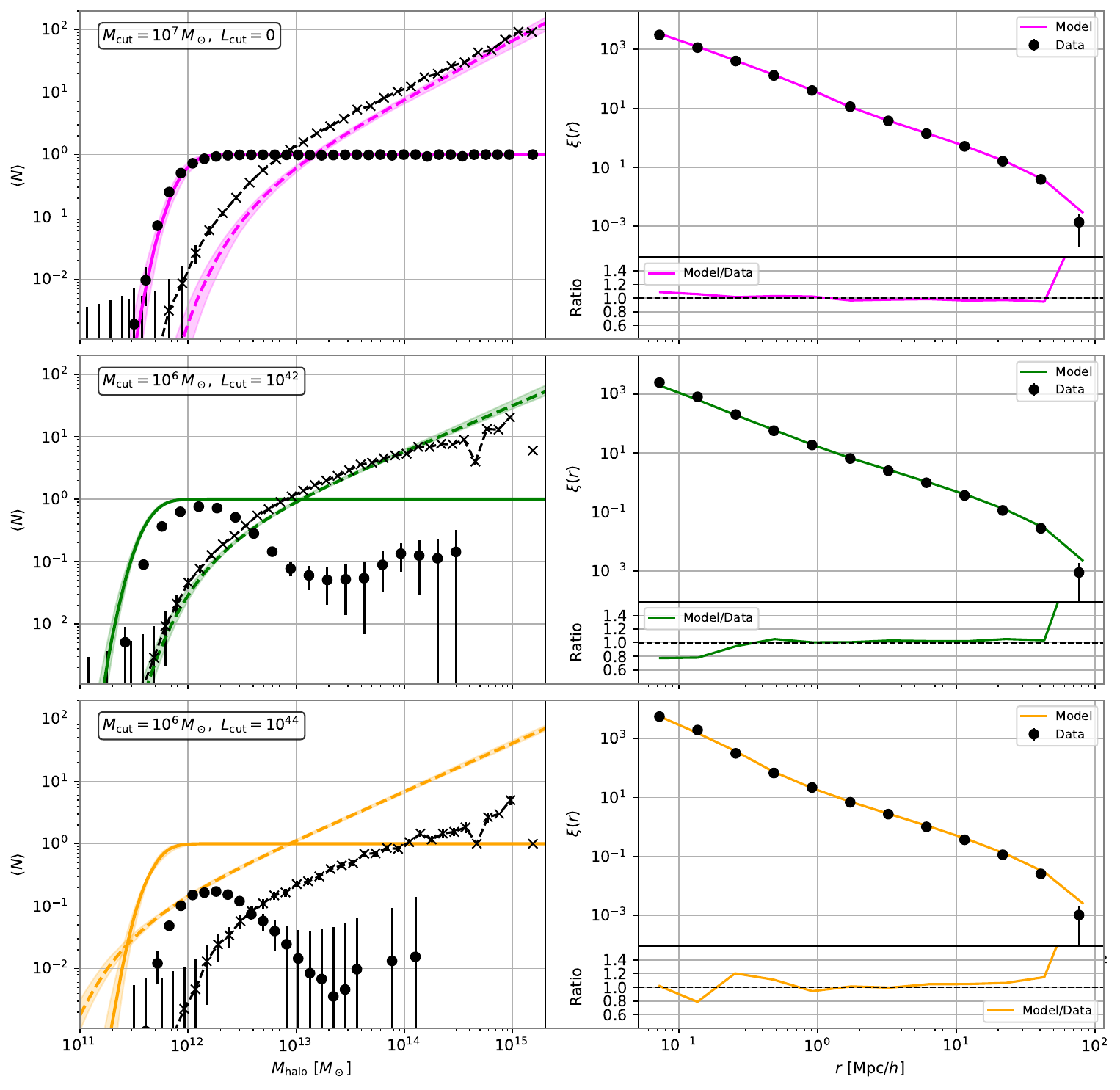}
  \caption{\textbf{Left:} Black dots (crosses) represent the directly measured Mean Occupation Function (MOF) of centrals (satellites) from the IllustrisTNG simulation (TNG300-1). Colored solid (dashed) lines show the median reconstructed MOF from the HOD fit described in Section~\ref{sec:clustering}, with shaded regions indicating the 16th and 84th percentiles. \textbf{Right:} Black dots show the measured $\xi(r)$ from TNG300-1, while colored lines represent the best-fit model predictions using the 5-parameter HOD. The lower panels display the ratio between the model and the simulation data.}
  \label{fig:xi}
\end{figure*}

\subsection{Two-Point Correlation Function}\label{sec:clustering}
As discussed, the two-point statistics, especially the two-point correlation function (2PCF), has been widely used along with the HOD framework to extract information about the AGN–halo connection. The real-space 2PCF of AGNs, denoted by $\xi(r)$, quantifies the excess probability, relative to a random distribution, of finding an AGN pair separated by a distance $r$ \citep{peebles80}. Within the halo model, this correlation function can be decomposed into two components: 
\[
\xi(r) = \xi_{\mathrm{1h}}(r) + \xi_{\mathrm{2h}}(r),
\]
representing intra-halo (1-halo) and inter-halo (2-halo) contributions, respectively.

The two-halo term dominates at large scales ($r \gtrsim 1\,\mathrm{Mpc}$) and accounts for correlations between AGNs residing in different halos. It can be approximated as \citep{b&w02}:
\begin{equation}
\xi_{\mathrm{2h}}(r) \approx 
\left[ 
n^{-1} \int_0^\infty \mathrm{d}M \, \frac{\mathrm{d}n}{\mathrm{d}M} \, \langle N(M) \rangle \, b_h(M)
\right]^2 \xi_m(r),
\end{equation}
where $n$ is the AGN number density, $\mathrm{d}n/\mathrm{d}M$ is the halo mass function, $\langle N(M) \rangle$ is the mean AGN occupation number for halos of mass $M$, $b_h(M)$ is the linear halo bias, and $\xi_m(r)$ is the matter correlation function. The expression in square brackets represents the effective large-scale bias of the AGN population.

At smaller scales, the one-halo term captures the contribution from AGN pairs within the same halo:
\begin{equation}
1 + \xi_{\mathrm{1h}}(r) \approx 
\frac{1}{4 \pi n^2 r^2} \int_0^\infty \mathrm{d}M \, \frac{\mathrm{d}n}{\mathrm{d}M} \, \langle N(N - 1) \rangle_M \, \frac{\mathrm{d}F_M(r)}{\mathrm{d}r},
\end{equation}
where $F_M(r)$ denotes the average cumulative fraction of AGN pairs within a halo of mass $M$ that are separated by less than $r$. This includes both central–satellite and satellite–satellite pairs and depends on the assumed spatial distribution of AGNs within halos, commonly modeled using NFW profiles \citep{zehavietal05}. However, recent work has highlighted the limitations of conventional HOD models, particularly in the context of the high-precision regime of upcoming galaxy surveys \citep{Mohrmann20,Boryana20}. Several extensions have been proposed to better capture the complexities of the galaxy–halo connection \citep{Boryana21,Adam22,Mohrmann23}.

In this work, we compute the real-space 2PCF for AGNs across different luminosity thresholds using \textsc{Corrfunc} \citep{Sinha20}, a publicly available Python package for computing clustering statistics. For our analysis, we employ the natural estimator \citep{d&p83}:
\begin{equation}\label{eqn:2pcf_estimator}
    \xi(r)=\frac{n_R}{n_D}\frac{DD(r)}{RR(r)} -1,
\end{equation}
which is a good estimator for periodic simulation boxes. We apply this analysis to the IllustrisTNG simulation (TNG300-1) at $z=0.5$ (intermediate between $z=1.0$ and $z=0.1$), using three samples: two luminosity-selected AGNs ($L > 10^{42}$ and $L > 10^{44}$ erg/s) and one mass-selected SMBH($M_{bh} > 10^{7}\, M_\odot$). We now employ the `true' correlation functions to extract the HODs and denote them as `reconstructed' HODs. For this we generate the mock AGN catalogs using \textsc{Halotools} \citep{Halotools}, populating dark matter halos from the Bolshoi-Planck dark-matter-only simulation. We note that the halos in bolplanck are identified using \textsc{Rockstar} while halos in TNG are FoF halos. Our results show mild differences in the halo mass function for these two simulations and we adopt a simple correction factor of $\sim 1.4$ between the two definitions of the halo mass to match the halo mass functions of the two simulations. We adopt a 5-parameter HOD model (see Eqs.~\ref{eqn:centralBH} and~\ref{eqn:satellite}) and employ MCMC sampling via the \textsc{emcee} package \citep{emcee13} to fit the `true' $\xi(r)$.  

Our results are shown in Figure~\ref{fig:xi}. The top panel shows the comparison for the mass-selected SMBH sample. As noted before, the MOF of mass-selected SMBH is well described by the 5 parameter model and hence the slight differences we observe in the satellite MOF reconstruction, is mostly dominated by the differences in the halo modeling between the two simulations. The middle and the bottom panels reveal significant discrepancies between the `reconstructed' (solid line) and the `true' (black data points) HODs, particularly for the most luminous AGN sample (bottom panel). When we compare these differences with the mass-selected sample, we note that the offsets are way beyond the differences in halo modeling in the two simulations. We discuss these results further in \S 4.

\section{Discussions and Conclusions} \label{Discussions}

Studies of AGN clustering have been the key tool in identifying the connection between AGN and their host dark matter halos \citep[e.g.,][]{costa24,shen09, hickoxetal09, shen13, myersetal06, eftekharzadehetal15,pizzatietal24a, pizzatietal24b}. The HOD modeling of clustering statistics, is an useful technique to extract the full host halo mass distributions of AGN. The popular HOD model for AGN occupation so far has been similar to the 5 parameter HOD models of galaxies \citep{zhengetal07, chatterjeeetal12}. In this work, using the TNG and the SIMBA simulations, we show that while the redshift independent 5-parameter HOD model is an adequate description of the mass-selected SMBH occupation (Fig.\ref{fig:Lum_Evolution}, left panel of Fig.\ref{fig:Redshift_Evolution_SMBH}, top panels of Fig.\ref{fig:TNG_v_SIMBA}, left column of Fig.\ref{fig:SIMBA_Feedback}), it does not explain the luminosity-selected AGN HOD. For a luminosity-selected AGN sample, the central occupation function strongly deviates from a step-function with a peak at a halo mass (log) scale of $11.5-12 M_{\odot}$ and a strong suppression at the mass scale of $13-13.5M_{\odot}$ (group and low mass cluster scale, Fig.\ref{fig:Lum_Evolution}) at all redshifts and over all luminosity bins. However, it is observed that the suppression is much stronger at low redshift and the occupation function follows a step-function like behavior as we go to higher redshifts (Fig.\ref{fig:Redshift_Evolution_SMBH} right panel).

Quasars representing the brightest AGN population have been distinctly used as a probe to understand the AGN-halo connection at high redshifts. From studies of quasar clustering, it is inferred that quasars preferentially reside at a typical halo mass scale ($\sim 10^{12.5} M_{\odot}$) at all redshifts with increasing bias \citep[e.g.,][]{costa24,bhowmicketal22,dimatteoetal12,shen09, hickoxetal09, shen13, myersetal06, eftekharzadehetal15,pizzatietal24a, pizzatietal24b}. It was understood that the phenomenon is directly involved with the gas cooling processes in halos \citep{Monaco07,Viola08,Best06,Cattaneo09}. Previously, the host halo mass distribution of quasars were extracted based on the simple 5-parameter HOD model \citep[e.g.,][]{richardsonetal12, mitraetal18, graysonetal23}. Similar methods have been utilized for X-ray selected AGN \citep[e.g.,][]{richardsonetal12, allevatoetal14, allevatoetal11}. Our study reveals that not only is the luminosity-selected AGN HOD different from the 5 parameter model, but also the functional form evolves with redshift (gradual convergence to the standard model at higher redshifts). As illustrated in \S 3.4, we note that this bias in modeling the HOD has a significant impact on extraction of host dark matter halos of AGN/quasars from clustering studies. 

When we compare our AGN MOF, with that of galaxies in the simulation, we find a similar suppression of the occupation fraction of star-forming galaxies at the characteristic halo mass scale (Fig.\ \ref{fig:Star_Formation_HOD}) similar to the luminosity-selected AGN HOD. Previously \citet{dave19} reported that star formation is significantly suppressed due to the jet mode of feedback in the SIMBA simulation runs. In addition, it has been observed that jet feedback in the SIMBA simulation has a very strong effect on observables such as the X-ray emission \citep{KarChowdhury22} and Sunyaev-Zeldovich signal \citep{Chakraborty23} in group-to-cluster scale halos. It is further seen that the kinetic mode (equivalent to jet mode) is the dominant mode of feedback between $z=0-2.0$ in TNG \citep{weinbergeretal18} and is responsible for quenching star formation.
Thus, it is likely that the strong mode of feedback from the AGN/quasar, combined with the lack of cold gas in larger halos, suppresses quasar activity and star formation simultaneously in group to low mass cluster sized halos.

While the central occupation is suppressed, the satellite occupation of luminosity-selected AGN increases with halo mass exhibiting a power-law like behavior and the fits to the power-law are shown in Fig.\ref{fig:ParameterVariation.pdf}. In Fig.\ref{fig:fsat}, we show the global satellite fraction of quasars (defined as our simulation sample with L$_{bol} > 10^{45}$ergs s$^{-1}$) and our mass-selected SMBH sample (defined as M$_{bh} > 10^{6}$M$_{\odot}$). As shown in Fig.\ref{fig:fsat}, the global satellite fraction increases at low redshifts ($z < 1.0$) suggesting that the nearby universe has higher numbers of quasars as satellites compared to centrals. While central quasars are associated with the gravitational potential of the host halo, satellite quasars are believed to be associated with secondary processes (e.g., galaxy mergers). But studies show that mergers are more frequent in the early Universe compared to the nearby one with the peak lying between $z=3.0$ and $z=1.0$ \citep[e.g.,][]{hopkinsetal06, boweretal06, bundyetal09, bharadwajetal18}. Thus the satellite fraction of quasars is opposite to that trend in the low redshift Universe. The plausible explanation of this trend could be linked to feedback effects from the central quasar. While the feedback from the central quasar inhibits its own growth, the satellites remain relatively unaffected within their own potential wells \citep{chatterjeeetal12, mitraetal18}, allowing a higher fraction of satellites to shine as bright quasars at lower redshifts. The situation is different for the mass-selected SMBH sample where the satellite fraction slowly increases from $z=4.0$ to $z=1.0$ and saturates to $\sim 25 \%$, more reflective of the merger rates in the Universe.

To compare our findings with observations we require to extract the HOD of quasars/AGN. While direct measurement of the occupation fraction is possible \citep{chatterjeeetal13}, it would be limited to lower redshifts and will be affected by the mass bias of host dark matter halos \citep{chatterjeeetal13}. Thus, the indirect route for extracting the quasar occupation function from the clustering measurements still provides the promising alternative \citep[e.g.,][]{shen13, richardsonetal12, mitraetal18, graysonetal23, Yuanetal24}. Such indirect method would require an apriori model of the HOD. Our results show that a significant bias is introduced by assuming an `inadequate' HOD to model the real space correlation function of AGNs. We find that reconstructing an HOD with an `inadequate' model could lead to understimating the host halo mass by a factor of $\sim4$. It is thus important to carefully quantify these theoretical biases while comparing with observed data \citep[e.g.,][]{desi24a}. Moreover, our studies reveal that the HOD is evolving significantly with redshift challenging the assumption of modest redshift evolution \citep[e.g.,][]{richardsonetal12}. Hence samples where quasar sources are stacked over a redshift bin for cosmological analysis will suffer theoretical biases when the redshift dependence of host halos are not taken into account. 

While both SIMBA and TNG qualitatively reveal similar results, it is important to note that the evolution of the HOD is sensitive to the parameters of the subgrid physics. It is thus suggestive to utilize a full parameter space study of the subgrid physics in cosmological simulations (namely accretion and feedback models) to characterize their effect on AGN and galaxy occupation functions and their relationship with the underlying dark matter halo population. Despite smaller boxsizes we propose using the CAMELS simulations \citep{CAMELS_presentation, CAMELS_DR1, CAMELS_DR2} to provide useful clues in this direction. Modeling of AGN accretion and feedback in cosmological simulations provides us the necessary theoretical template, for interpreting the underlying connection between AGN and large scale structure, from  current and upcoming survey datasets. Our study provides a context in that direction and confirms that the AGN/quasar HOD, obtained from hydrodynamical simulations, is evolving and is not fully consistent with the simple 5 parameter HOD model. It is thus important to carefully analyze the impact of feedback and other subgrid physics in interpreting large scale structure observables using AGN/quasar surveys. 

We now summarize our main findings below. 
\begin{itemize}
    \item Mass-selected SMBH samples constructed from hydrodynamic simulations show weak redshift evolution in their halo occupation functions suggesting minimal redshift and environmental dependence in black hole growth. The mean occupation function of black holes is well modeled by a softened step function with a power-law as has been seen in previous studies.
    \item When black holes are selected by their luminosities, the central occupation function departs from a  simple softened step function model. There exists a characteristic halo mass scale ($ \sim 10^{13}M_{\odot}$) where the central occupation is suppressed at all redshifts. The suppression is significantly stronger at lower redshifts. Similar to AGN, there is evidence of suppression of the number of star-forming galaxies at a similar characteristic halo mass scale. While more studies are needed, it is understood that the lack of availability of cold gas and the onset of strong feedback from the AGN are responsible for the suppression of AGN activity as well as star-formation at these mass scales. 
\item The redshift evolution of the mean occupation function is strongest from z=0.1 to z=1 for the luminosity-selected AGNs. Central occupation function evolves substantially over redshift, where, at higher redshifts the step-function behavior is largely observed.
    
    \item  While the central occupation deviates from the usual model, the satellite occupation still shows a power-law like behavior. The satellite numbers of luminosity-selected AGN samples exhibit a Poisson like distribution for different halo mass bins, similar to what has been observed in previous studies. The departure from a Pure Poisson distribution is predominantly sub-Poissonian in nature. 
     The satellite population function evolves weakly with redshift, while the global satellite fraction of quasars exhibits strong evolution with redshift. The satellite fraction increases substantially at low redshifts, consistent with the findings of clustering based studies. 
     
     \item The evolving HOD has significant effect on clustering analysis. Our results show that using an inadequate HOD model to infer the SMBH–halo connection from the real-space two-point correlation function results in an underestimation of the typical host halo mass by a factor of $\sim4$ and describes a completely different distribution of host halo masses while overestimating or underestimating the satellite fraction. In addition, our results show that samples of AGN, constructed over a broad redshift range is likely to add theoretical biases in extracting cosmological information. 

    \item While there is broad agreement in the overall relationship between dark matter halos and AGN/quasars in cosmological simulations, resolution effects and modeling of the subgrid physics could still be an issue in physically interpreting the results and hence appropriate caution must be taken while comparing the results with observations. However, despite these limitations, cosmological simulations do provide the appropriate machinery to interpret the co-evolution paradigm of quasars and future studies should aim at calibrating the subgrid physics parameters (namely feedback and accretion models) to truly comprehend the physical effects of AGN growth and feedback in the cosmological landscape.

\end{itemize}

\section*{Acknowledgements}
 SC acknowledges financial support from ANRF through the POWER Fellowship (SPF/2022/000084) and the CRG/2020/002064 grant. AC acknowledges financial support from UGC through UGC-NET Junior Research Fellowship. AC and SC thank Shadab Alam, Saumyadip Samui, Aseem Paranjape, and Supratik Pal for useful discussions in directing the analysis of the paper and Ritaban Chatterjee for useful discussions and help with providing computing facilities. SC and AC particularly thank the anonymous referee for pointing to many references and providing useful suggestions, which greatly helped in improvement of the draft. SC also thanks Arnatri Samajdar, Anwesh Majumdar and Kaustav Mitra, for some aspects of the analysis at a preliminary level and the Inter University Center for Astronomy and Astrophysics for usage of their facilities through the IUCAA associateship program.

\bibliography{mybib}{}
\bibliographystyle{aasjournal}

\appendix 
\section{Satellite Distribution}
\begin{figure}[h] 
    \includegraphics[width=0.44\textwidth]{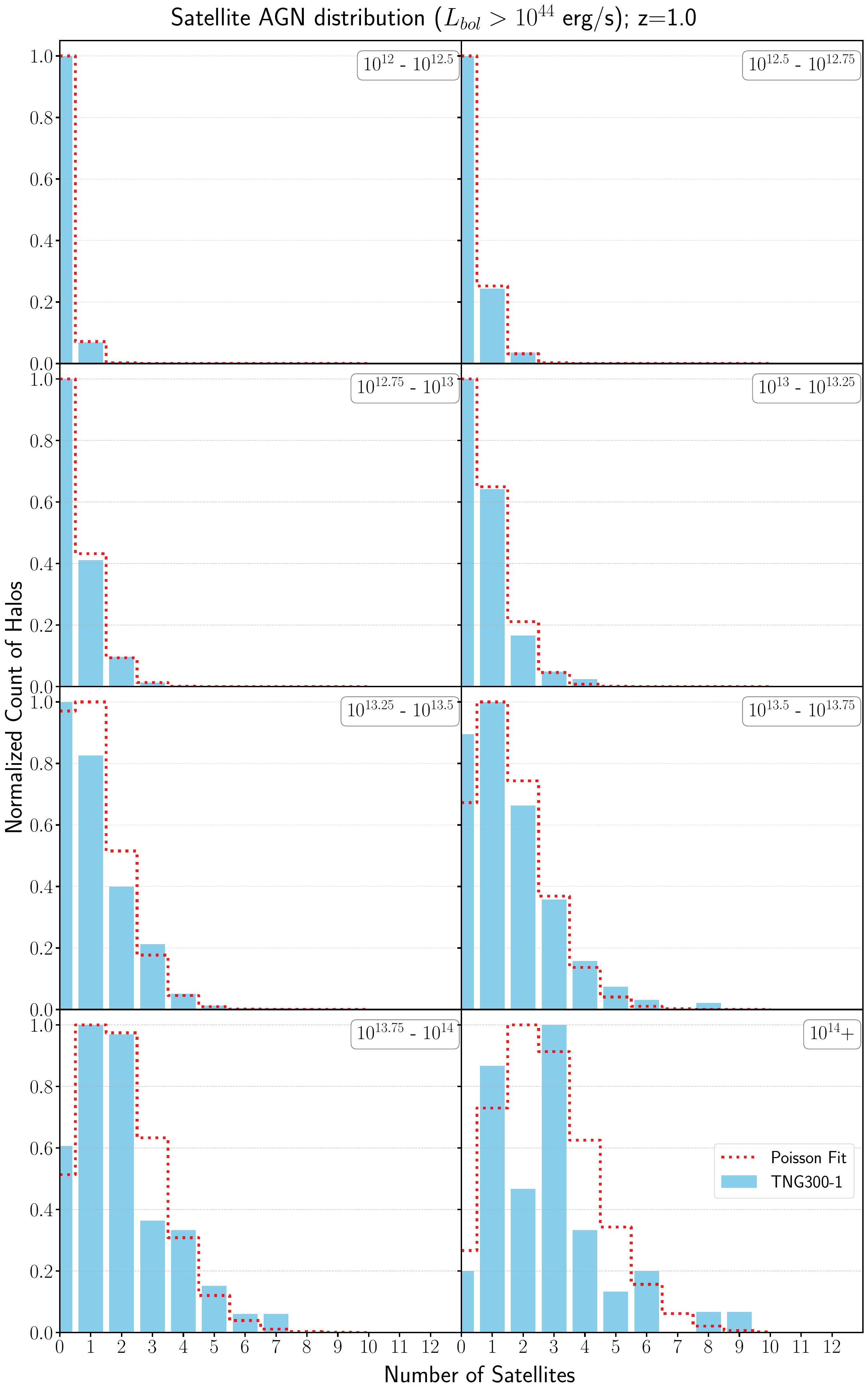}
        \caption{Normalized number distributions of satellite AGNs ($L_{bol}>10^{44}erg/s$) in host halo mass bins. The blue bar plots show the actual distribution of AGNs in halos (normalized to the maximum number), and the red dashed histograms depict the theoretical Poisson distribution with the sample mean in the particular mass bin. The number zero represents the number of halos in the mass bin that do not host any AGN ($L_{bol}>10^{44}erg/s$). The number distribution is close to a Poisson distribution, justifying the Poisson error bars. The departure from Poisson is mostly sub-Poissonian in nature.}
    \label{fig:poisson_fitted_plot.pdf}
        \end{figure}

     While using \textsc{HaloTools}, we assumed the satellites to be poissonian. In Fig. \ref{fig:poisson_fitted_plot.pdf} we show the normalized number distribution of satellite AGNs ($L_{bol}>10^{44}erg/s$) in different host halo mass bins. The blue histograms represent the satellite number distributions, while the red dotted line is the theoretical Poisson distribution generated from the sample mean. As observed, the number distribution is close to a Poisson distribution in all halo mass bins \citep{chatterjeeetal12, degrafetal11b} with departures that exhibit more sub-Poissonian behavior.

\section {Resolution Effects}

\begin{figure*}[t]
 \begin{center}
    \includegraphics[width=0.7\textwidth]{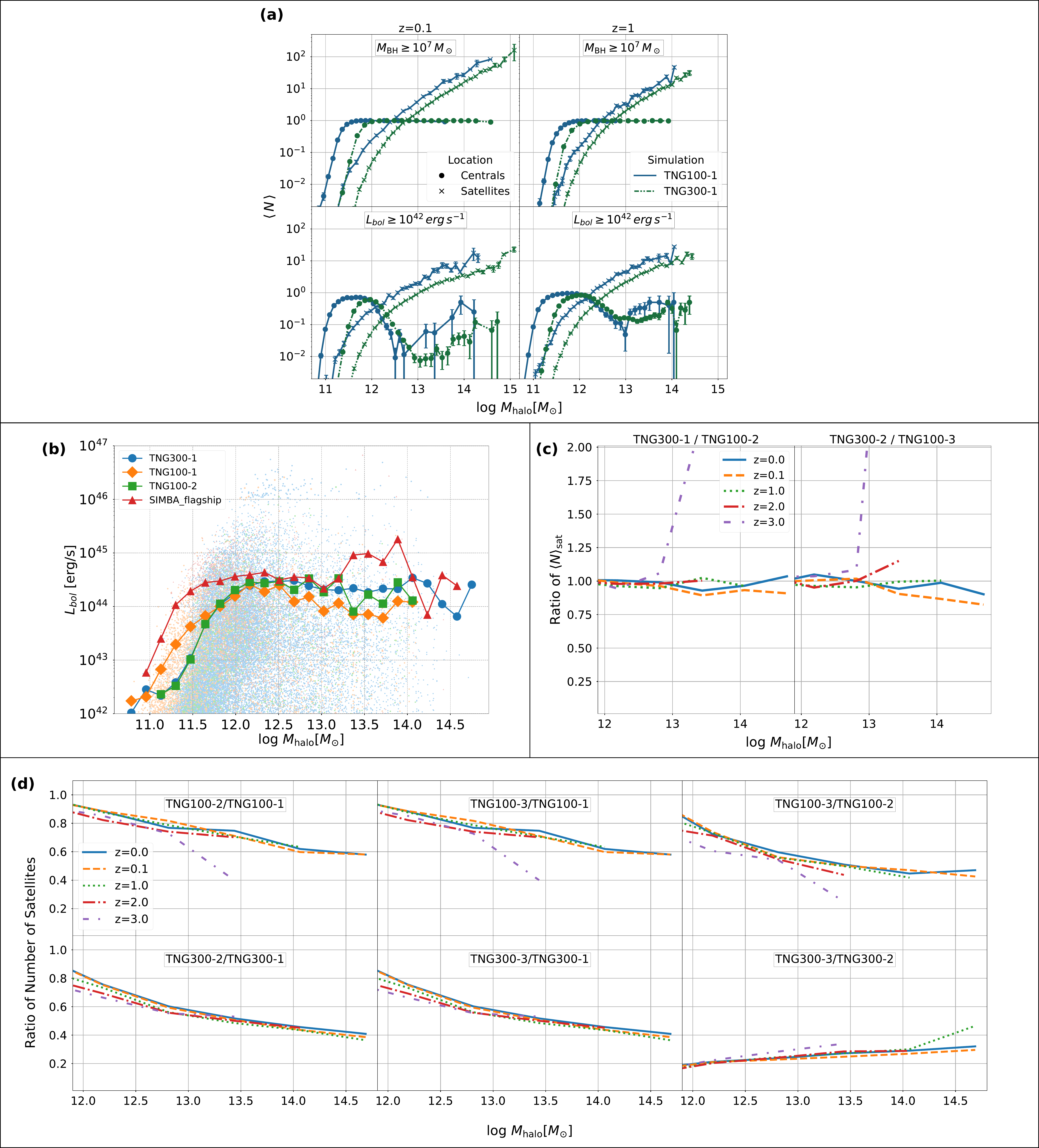} 
    \caption{{\bf (a) :} Comparison between the Mean Occupation Function of AGN in TNG100-1 (blue) and TNG300-1 (green) at redshifts $z = 0.1$ (left) and $z = 1$ (right). Dots and crosses represent centrals and satellites respectively. \textbf{Top :} Mass-selected SMBH with $M_{BH}>10^{7}\,M_{\odot}$. \textbf{Bottom :} Luminosity-selected AGN with $L_{\text{bol}} > 10^{44} \, \text{erg} \, \text{s}^{-1}$. The error bars on data points are $1\sigma$ Poisson error bars. While both in the SMBH and AGN population the central occupation functions match, there are significant differences in the satellite population. {\bf (b) :} AGN($L_{bol}\,\geq\,10^{42}\,erg/s$) luminosity versus host halo mass ($20 \%$ of the data points are randomly plotted for visualization purpose). The blue, orange, green, and red points are for TNG300-1, TNG100-1, TNG100-2, and SIMBA$_{flagship}$ respectively at z=1. The solid lines represent the average AGN luminosity at a given halo mass. {\bf (c) :}: Ratio of Mean Occupation Function (MOF) of satellite SMBH in simulations with same resolution but different boxsize. \textbf{Left} : Ratio of MOF between TNG300-1 and TNG100-2. \textbf{Right} : Ratio of MOF between TNG300-2 and TNG100-3. {\bf (d) :} Ratio of the number of satellite supermassive black holes (SMBH) as a function of halo mass, comparing simulations with varying resolutions. IllustrisTNG simulations with a 100 Mpc box size and IllustrisTNG simulations with a 300 Mpc box size.} 
    \label{fig:appendix_plots.pdf}
     \end{center}
\end{figure*}

  In cosmological simulations, black holes are modeled with sub-resolution prescriptions for seeding, growth, and its interaction with surrounding gas. We test for the resolution dependence of the MOF of AGN by comparing TNG100-1 and TNG300-1 at redshifts $z = 0.1$ and $z = 1.0$ (panel (a) of Fig. \ref{fig:appendix_plots.pdf}). We see that both TNG100-1 and TNG300-1 produce similar MOFs with a dip in the central AGN occupation at a characteristic mass scale. However, TNG100 consistently shows higher MOF values for all mass bins particularly for the satellites. The effect is more prominent, especially in halos with $M_{\text{halo}} < 10^{12.5} \, M_{\odot}$. As TNG100-1 has a higher resolution, it identifies smaller AGN better than TNG300-1, producing a better representation of the AGN occupation. 
 
 The convergence between TNG100-1 and TNG300-1 improves at higher halo mass, especially for the central AGN at $z = 1.0$. While TNG-100 better resolves smaller mass halos, it has less statistics in the high mass end due to its smaller box size. In panel (b) of Fig.\ref{fig:appendix_plots.pdf} we plot AGN luminosity ($\geq 10^{42}$ ergs/s) with their host halo mass for different simulations at z=1. We see that the TNG300-1 and TNG100-2 with similar resolutions produce similar luminosity distributions. However, TNG100-1, with a higher resolution than TNG300-1 and TNG100-2 has lower average AGN luminosity at halos above $10^{12}\,M{\odot}$ and higher average luminosity below $10^{12}\,M{\odot}$. SIMBA on the other hand produces more luminous AGNs in every halo mass bin. Solid lines represent the average AGN luminosity in a given halo mass bin. To do a more comprehensive study of the effect of resolutions, we perform the analyses for similar box sizes with varied resolutions and similar resolutions with varied box sizes for the TNG-300 and TNG-100 simulations. We note that when the resolutions are similar we observe high convergence between halo and black hole numbers over a wide range of redshifts. In panel (c) of Fig. \ref{fig:appendix_plots.pdf} we compare simulations of different box sizes with identical resolution to test for convergence. We find that the average number of satellite SMBH are consistently same in the two simulation volumes of a given resolution.
The resolution effect is evident when we compare multiple realizations of the IllustrisTNG suite for effects of resolution, we compare the number of satellite mass-selected SMBHs identified in each simulation. In panel(d) of Fig. \ref{fig:appendix_plots.pdf}, we compare simulations of the same boxsize but different resolution. We find that simulations with higher resolution predict higher number of satellite SMBH. The discrepancy is stronger in more massive halos suggesting stronger effects of resolution in denser environments. 

\end{document}